\documentclass[10pt]{extarticle}
\usepackage{graphicx}
\usepackage{amssymb,amsmath}
\def\beq{\begin{eqnarray}}
\def\eeq{\end{eqnarray}}
\usepackage[titletoc,toc,title]{appendix}

\usepackage{amsfonts}

\begin{document}

\title{On the calculation of exact sum rules of rational order for quantum billiards}
\author{Paolo Amore \\
\small Facultad de Ciencias, CUICBAS, Universidad de Colima,\\
\small Bernal D\'{i}az del Castillo 340, Colima, Colima, Mexico \\
\small paolo@ucol.mx}

\maketitle

\begin{abstract}
	We study the Helmholtz equation for a heterogeneous system in $d$ dimensions and show that it is 
	possible to calculate exactly the sum rules of rational order using perturbation theory
	by relating the sum rules to suitable traces.
	The present calculation is restricted to a positive definite spectrum, e.g. Dirichlet boundary conditions, 
	and it reproduces the result previously obtained in Ref.~\cite{Amore12}, working directly
	with Rayleigh-Schr\"odinger perturbation theory.
\end{abstract}

\maketitle

\section{Introduction}
\label{intro}

Spectral zeta functions are defined as
\begin{equation}
Z(s) \equiv \sum_n \frac{1}{E_n^s} \label{eQ_intro_1}
\end{equation}
where $E_n$ are the eigenvalues of the Hamiltonian operator and $s>s_0$ is a real exponent,  $s_0$  being 
the minimal value of $s$ for which the series (\ref{eQ_intro_1}) converges.

In this paper, we consider the Helmholtz equation for a heterogeneous system in $d$ dimensions, i.e.,
\begin{equation}
- \Delta_d \Psi_n({\bf x}) = E_n \Sigma({\bf x}) \Psi({\bf x}) \hskip 0.2cm , \hskip 0.2cm {\bf x} \in \Omega
\label{eQ_intro_2}
\end{equation}
where $\Delta_d \equiv \partial^2/\partial x_1^2+ \dots + \partial^2/\partial x_d^2$ is the Laplacian operator, $\Sigma({\bf x}) >0$ is the density, $\Omega$ is a d-dimensional region where an orthonormal 
and complete set of eigenfunctions of the negative Laplacian is known.

In general it is not possible to calculate $Z(s)$ exactly for arbitrary values of $s$, since the exact eigenvalues 
appearing in eq.~(\ref{eQ_intro_2}) are not known. An exception to this rule occurs when $s$ takes integer values,
$s>s_0$; in this special case the sum rule can be obtained directly in terms of a trace, exploiting the invariance 
of the trace with respect to unitary transformations and the completeness of the basis of the eigenfunctions
of the negative Laplacian.

In this way, Itzykson, Moussa and Luck were able to obtain in Ref.~\cite{Itzykson86} the sum rules for an arbitrary two dimensional  region with Dirichlet boundary conditions at the border~\footnote{In this case a conformal transformation can be used to map the original region to a disk and the density in eq.~(\ref{eQ_intro_1}) is related to the conformal map.}.
We mention Refs.~\cite{Berry86,Steiner85,Steiner87, Kvitsinsky96, Dittmar02,Dittmar11,Dostanic11}
as an example of works concerning sum rules (further references can be found by the reader in those articles).

Using  a similar approach, in refs.~\cite{Amore13A,Amore13B,Amore14,Amore18,Amore19} 
we have derived general integral expressions for the spectral sum rules of inhomogenous strings and membranes, 
for different boundary conditions. In particular Refs.~\cite{Amore14,Amore19} deal with the special case of a spectrum
containing a null eigenvalue, for which the traces are in principle ill--defined and a suitable renormalization has to be performed.

It is not straightforward to relate the sum rule  to a trace in the case of non--integer exponents, on the other hand: for this reason, in Ref.~\cite{Amore12}, we followed a different approach, which exploits 
the Rayleigh-Schr\"odinger perturbation theory, expressing the sum rule in eq.~(\ref{eQ_intro_1}) directly in term
of the perturbative expressions for the eigenvalues. Eq.~(9) of that paper provides an exact expression for
the sum rule $Z(s)$, correct up to second order in the density.

The purpose of this paper is to confirm eq.(9) of Ref.~\cite{Amore12} without resorting to Rayleigh-Schr\"odinger perturbation theory and working directly with suitable traces, for cases where the spectrum is positive definite, e.g. Dirichlet boundary conditions.

The paper is organized as follows: in section \ref{sec:GF} we introduce Green's function of order $1/2$ and 
we use perturbation theory to calculate the first few orders (up to order three explicitly and up to order eight,
reported in the first appendix, implicitly); in section \ref{sec:GF_rational}, we extend the previous results and
introduce Green's functions of order $1/N$ ($N=2,3,\dots$) and use perturbation theory to calculate their general expressions up to second order; in section \ref{sec:sumrules} we obtain the sum rules of rational exponent in term of the traces involving products of Green's function of suitable orders and reproduce eq.(9) of Ref.~\cite{Amore12}. Finally in section \ref{sec:concl} we  draw our conclusions. The perturbative expressions for the coefficients of the spectral decomposition of the Green's  function of order $1/2$ are reported in appendix \ref{appA}; in appendix \ref{appB} we report special values for the  functions $\Delta$, $\eta$ and $\xi$, introduced in section \ref{sec:GF_rational}.

\section{Green's functions of order $1/2$}
\label{sec:GF}

To begin our discussion, let us define
\begin{equation}
G(x,y) \equiv \sqrt{\Sigma(x)} G_0(x,y) \sqrt{\Sigma(x)} \ , 
\end{equation}
where $\Sigma(x) >0$ is the density and $G_0(x,y)$ is the Green's function of the homogeneous problem.
We can convince ourselves that $G(x,y)$ is the Green's function associated to the operator $\hat{O} \equiv \frac{1}{\sqrt{\Sigma(x)}} (-\Delta) \frac{1}{\sqrt{\Sigma(x)}}$ since
\begin{equation}
\hat{O} G(x,y) =  \frac{1}{\sqrt{\Sigma(x)}} (-\Delta) G_0(x,y)  {\sqrt{\Sigma(y)}}  = \delta(x-y) \ .
\end{equation}

Next introduce a new function, $\tilde{G}(x,y)$, satisfying the property:
\begin{equation}
\int \tilde{G}(x,z)\tilde{G}(z,y) dz  = G(x,y) \ .
\label{GF_eQ_1}
\end{equation}

We can easily see that eq.~(\ref{GF_eQ_1}) is satisfied provided that %$\tilde{G}(x,y)$ complies with the property
\begin{equation}
\hat{O}^{1/2} \tilde{G}(x,y) = \delta(x-y) \ . 
\end{equation}

As a matter of fact, we have
\begin{equation}
\hat{O} G(x,y) = \hat{O}^{1/2} \int \left[\hat{O}^{1/2}\tilde{G}(x,z) \right] \tilde{G}(z,y) dz =
\hat{O}^{1/2}  \tilde{G}(x,y) =  \delta(x-y) \ .
\end{equation}

Let us now decompose $\tilde{G}$ in the basis of the {\sl homogeneous} problem:
\begin{equation}
\begin{split}
\tilde{G}(x,y) &= \sum_{n,m} q_{nm} \psi_n(x)  \psi_m^\star(y) \ , 
\end{split}
\end{equation}
where
\begin{equation}
q_{nm} = \int \int \psi_n^\star(x) \tilde{G}(x,y) \psi_m(y) dxdy \ .
\end{equation}

Using this expression in the left--hand side of eq.~(\ref{GF_eQ_1}), we have
\begin{equation}
\begin{split}
\int \tilde{G}(x,z)\tilde{G}(z,y) dz % &= \sum_{n,m} \sum_{n',m'} Q_{nm} Q_{n' m'} \psi_n(x) \psi_{m'}(y) \delta_{m,n'} \nonumber \\
&= \sum_{n,m,m'} q_{nm} q_{n' m'} \psi_n(x) \psi_{m'}(y) \ .
\label{GF_eQ_2a}
\end{split}
\end{equation}

Similarly, the right hand side of eq.~(\ref{GF_eQ_1}) becomes
\begin{equation}
\begin{split}
G(x,y) &= \sum_{n,m} \left[\sum_r \frac{\langle n| \sqrt{\Sigma}| r \rangle \langle r| \sqrt{\Sigma}| m \rangle}{\epsilon_r} \right] \ \psi_n(x) \psi_m^\star(y) \\
&= \sum_{n,m} Q_{nm} \ \psi_n(x) \psi_m^\star(y) \ ,
\label{GF_eQ_2b}
\end{split}
\end{equation}
where $Q_{nm}$ are the coefficients of the spectral expansion of $G(x,y)$. \\
By equating eqs.~(\ref{GF_eQ_2a}) and (\ref{GF_eQ_2b}) we finally obtain the matrix equation
\begin{equation}
\sum_{r} q_{nr} q_{r m} = \sum_r \frac{\langle n| \sqrt{\Sigma}| r \rangle \langle r| \sqrt{\Sigma}| m \rangle}{\epsilon_r} \ .
\label{GF_eQ_3}
\end{equation}

The exact solution of eq.~(\ref{GF_eQ_3}), which would provide the exact expression for the Green's function of order $1/2$ cannot be obtained for an arbitrary density $\Sigma$, and therefore it is convenient to resort to perturbation theory. In this case we assume a mild inhomogeneity and write
\begin{equation}
\Sigma(x) =  1 + \lambda \sigma(x) \ ,
\end{equation}
with $|\sigma(x)| \ll 1$ for all $x$. $\lambda$ is a power--counting parameter that will be set
to $1$ at the end of the calculation.

Therefore
\begin{equation}
\sqrt{\Sigma(x)} = \sum_{j=0}^\infty \left( \begin{array}{c}
1/2 \\
j \\
\end{array} \right) \lambda^j \sigma(x)^j \ .
\end{equation}

Similarly, we write
\begin{equation}
\begin{split}
q_{nm} &= \sum_{j=0}^\infty q_{nm}^{(j)} \lambda^j \ , \\
Q_{nm} &= \sum_{j=0}^\infty Q_{nm}^{(j)} \lambda^j \ . \\
\end{split}
\end{equation}

It is straighforward to see that 
\begin{eqnarray}
Q_{nm}^{(k)} = \sum_{j=0}^k \left( \begin{array}{c}
1/2 \\
j \\
\end{array} \right)  \ \left( \begin{array}{c}
1/2 \\
k -j \\
\end{array} \right)  \sum_r \frac{\langle n| \sigma^j| r \rangle \langle r| \sigma^{k-j}| m \rangle}{\epsilon_r} \ .
\label{eQ_qk}
\end{eqnarray}

In particular the first three coefficients read
\begin{equation}
\begin{split}
Q_{nm}^{(0)} &= \frac{\delta_{nm}}{\epsilon_r} \ , \\
Q_{nm}^{(1)} &= \frac{1}{2} \left( \frac{1}{\epsilon_n}+ \frac{1}{\epsilon_m} \right) \ \langle n |\sigma|m \rangle \  ,  \\
Q_{nm}^{(2)} &= \frac{1}{4} \sum_r \frac{\langle n | \sigma | r\rangle \langle r | \sigma | m\rangle }{\epsilon_r}
-\frac{1}{8} \left( \frac{1}{\epsilon_n}+ \frac{1}{\epsilon_m} \right) \ \langle n |\sigma^2|m \rangle \ .
\end{split}
\end{equation}

By inserting eq.~(\ref{eQ_qk}) inside eq.~(\ref{GF_eQ_3}) and selecting the term of order $\lambda^k$,
we obtain the matrix equation
\begin{eqnarray}
\sum_{j=0}^k q_{nr}^{(j)}  q_{rm}^{(k-j)} = Q_{nm}^{(k)} \ .
%\sum_{j=0}^k \left( \begin{array}{c}
%1/2 \\
%j \\
%\end{array} \right)  \ \left( \begin{array}{c}
%1/2 \\
%k -j \\
%\end{array} \right)  \sum_r \frac{\langle n| \sigma^j| r \rangle \langle r| \sigma^{k-j}| m \rangle}{\epsilon_r} 
\label{GF_eQ_4}
\end{eqnarray}

The solutions of eqs.~(\ref{GF_eQ_4}) can be obtained iteratively starting from the lowest order ($k=0$) and moving to higher orders.

Before we proceed to the calculation of the first few orders, it is convenient to define:
\begin{equation}
\begin{split}
\Delta_{nm} & \equiv \frac{\frac{1}{\epsilon_n}+\frac{1}{\epsilon_m}}{\frac{1}{\sqrt{\epsilon_n}}+\frac{1}{\sqrt{\epsilon_m}}} \ , \\
\eta_{nm} & \equiv \frac{1}{\sqrt{\epsilon_n}}+\frac{1}{\sqrt{\epsilon_m}}  \ .
\end{split}
\label{eQ_delta_eta}
\end{equation}

The solution of order $0$  can be cast in the form 
\begin{equation}
q_{nm}^{(0)}  =  \frac{\delta_{nm}}{\sqrt{\epsilon_n}} =  \delta_{nm} \Delta_{nm}
\end{equation}
using eq.~(\ref{eQ_delta_eta}).

Similarly, the solutions up to third order can be obtained straightforwardly. They read
\begin{equation}
\begin{split}
q_{nm}^{(1)}  &=  \frac{1}{2} \ \frac{\frac{1}{\epsilon_n}+\frac{1}{\epsilon_m}}{\frac{1}{\sqrt{\epsilon_n}}+\frac{1}{\sqrt{\epsilon_m}}}
\langle n | \sigma | m \rangle = \frac{1}{2} \Delta_{nm} \langle n | \sigma | m \rangle  \ , \\
q_{nm}^{(2)}  &=  - \frac{1}{8} \Delta_{nm} \langle n | \sigma^2 | m \rangle + 
\frac{1}{4\eta_{nm}} \sum_r \left( \frac{1}{\epsilon_r} - \Delta_{nr} \Delta_{rm}\right) 
\langle n | \sigma | r \rangle \langle r | \sigma | m \rangle \  , \\
q_{nm}^{(3)}  &=  \frac{1}{16} \Delta_{nm} \langle n | \sigma^3 | m \rangle  \\
& -
\frac{1}{16 \eta_{nm}} \sum_r \left( \frac{1}{\epsilon_r} - \Delta_{nr} \Delta_{rm}\right) 
\left( 
\langle n | \sigma | r \rangle \langle r | \sigma^2 | m \rangle +
\langle n | \sigma^2 | r \rangle \langle r | \sigma | m \rangle \right)  \\
& - \sum_{r,s} \frac{\Delta_{nr}}{8 \eta_{nm} \eta_{rm}}
\left( \frac{1}{\epsilon_r} - \Delta_{rs} \Delta_{sm}\right) 
\langle n | \sigma | r \rangle \langle r | \sigma | s \rangle \langle s | \sigma | m \rangle \\
& - \sum_{r,s} \frac{\Delta_{rm}}{8 \eta_{nm} \eta_{nr}}
\left( \frac{1}{\epsilon_r} - \Delta_{ns} \Delta_{sr}\right) 
\langle n | \sigma | s \rangle \langle s | \sigma | r \rangle \langle r | \sigma | m \rangle \ .
\end{split}
\end{equation}

In Appendix \ref{appA} we report the perturbative corrections to the coefficients $q_{nm}$ up to order eight 
(in an implicit form). Observing  that the first term to order $k$ has the general form
$\left( \begin{array}{c}
\frac{1}{2} \\
k \\
\end{array}\right) \Delta_{nm} \langle n | \sigma^k | m \rangle$, we can perform a resummation obtaining 
\begin{equation}
q_{nm} \approx \sum_{k=0}^\infty \left( \begin{array}{c}
\frac{1}{2} \\
k \\
\end{array}\right) \Delta_{nm} \langle n | \sigma^k | m \rangle = \Delta_{nm} \langle n | \sqrt{\Sigma}| m \rangle \ .
\end{equation}

\section{Green's functions of order $1/N$}
\label{sec:GF_rational}

After having introduced the Green's functions of order $1/2$, we may generalize our discussion to the case
of Green's functions of order $1/N$, with $N$ positive integer . 

Now define $\tilde{G}_{[1/N]}(x,y)$ satisfying the property:
\begin{equation}
\int \tilde{G}_{[1/N]}(x,z_1) \tilde{G}_{[1/N]}(z_1,z_2) \dots \tilde{G}_{[1/N]}(z_N,y)  dz_1 \dots dz_N  = G(x,y)
\label{GFrat_eQ_1}
\end{equation}
with $N \geq 2$ (note: $N=2$ is a special case and has previously been considered).

We can decompose $\tilde{G}_{[1/N]}$ in the basis of the {\sl homogeneous} problem:
\begin{equation}
\begin{split}
\tilde{G}_{[1/N]}(x,y) &= \sum_{n,m} q_{nm}^{[1/N]} \psi_n(x)  \psi_m^\star(y) \ ,
\label{GFrat_eQ_2}
\end{split}
\end{equation}
where
\begin{equation}
q_{nm}^{[1/N]} = \int \int \psi_n^\star(x) \tilde{G}_{[1/N]}(x,y) \psi_m(y) dxdy \ .
\end{equation}

In analogy to what we have done before, we can use eq.~(\ref{GFrat_eQ_2}) inside eq.~(\ref{GFrat_eQ_1}) to obtain the  matrix equation
\begin{equation}
\sum_{r_1, \dots, r_N} q_{nr_1}^{[1/N]}  q_{r_1 r_2}^{[1/N]} \dots q_{r_N m}^{[1/N]}  = \sum_r \frac{\langle n| \sqrt{\Sigma}| r \rangle \langle r| \sqrt{\Sigma}| m \rangle}{\epsilon_r} \ .
\label{GFrat_eQ_3}
\end{equation}
From this point  we will avoid the superscript $[1/N]$ in the coefficients whenever possible.

We assume that the inhomogeneity is a perturbation and express the coefficients as a power series  as
\begin{equation}
q_{nm} = \sum_{j=0}^\infty q_{nm}^{(j)} \lambda^j \ .
\end{equation}

We generalize the previous definitions by introducing
\begin{equation}
\begin{split}
\Delta_{nm}^{[1/N]}  &\equiv \frac{\left( \frac{1}{\epsilon_n} + \frac{1}{\epsilon_m} \right)}{\sum_{j=0}^{N-1} \frac{1}{\epsilon_n^{(N-1-j)/N} \epsilon_m^{j/N}}}  \ , \\
\eta_{nm}^{[1/N]}  &\equiv \sum_{j=0}^{N-1} \frac{1}{\epsilon_n^{(N-1-j)/N} \epsilon_m^{j/N}}  \ ,
\end{split}
\end{equation}
which reduce to eqs.~(\ref{eQ_delta_eta}) when $N=2$.

The solutions of eq.~(\ref{GFrat_eQ_3}) up to second order take the form
\begin{equation}
\begin{split}
q_{nm}^{(0)}  &=  \frac{\delta_{nm}}{(\epsilon_n)^{1/N}} = \frac{N}{2} \Delta_{nm}^{[1/N]} \delta_{nm} \ , \\
q_{nm}^{(1)}  &= \frac{1}{2} \Delta_{nm}^{[1/N]} \langle n |\sigma|m \rangle \ , \\
q_{nm}^{(2)} &=  - \frac{1}{8} \Delta_{nm}^{[1/N]} \langle n | \sigma^2 | m \rangle   \\
&+ \frac{1}{4 \eta_{nm}^{[1/N]}} \sum_r \langle n | \sigma | r \rangle \langle r | \sigma | m \rangle \left( \frac{1}{\epsilon_r}
-\Delta_{nr}^{[1/N]} \Delta_{rm}^{[1/N]}  \xi_{nrm}^{[1/N]}\right) \ ,
\end{split}
\end{equation}
where we have defined
\begin{equation}
\xi_{nrm}^{[1/N]} \equiv  \sum_{j=0}^{N-2} \sum_{l=0}^{N-2-j} \frac{1}{\epsilon_n^{j/N} \epsilon_m^{(N-2-j-l)/N} \epsilon_r^{l/N}}\ .
\end{equation}

The explicit expressions of $\Delta_{nm}^{[1/N]}$, $\eta_{nm}^{[1/N]}$ and $\xi_{nrm}^{[1/N]}$ for $N=1,\dots, 4$ are reported in Appendix \ref{appB}.

\section{Sum rules of rational order}
\label{sec:sumrules}

The results obtained in the previous sections  allow us to obtain explicit expressions for the sum rules
of rational order by expressing them in terms of the traces involving the coefficients of the spectral decomposition
of the Green's functions of rational order. Next we will discuss the cases of sum rules of order $1+1/N$ and $1/N+1/N'$, with $N$ and $N'$ integers ($N,N' \geq 2$) and prove that for the two cases we obtain the {\sl same} functional form.

\subsection{Order $1+1/N$}

The sum rule of order $1 +1/N$ takes the form
\begin{equation}
\begin{split}
Z\left(1+\frac{1}{N}\right) &= \sum_{n,r} Q_{nr} q_{rn}^{[1/N]}\\
&= \sum_{n,r} Q_{nr}^{(0)} q_{rn}^{[1/N] (0)} \\
&+ \lambda \sum_{n,r} \left[ Q_{nr}^{(0)} q_{rn}^{[1/N] (1)} +  Q_{nr}^{(1)} q_{rn}^{[1/N] (0)}  \right] \\
& + \lambda^2 \sum_{n,r}\left[ Q_{nr}^{(1)} q_{rn}^{[1/N] (1)} + Q_{nr}^{(2)} q_{rn}^{[1/N] (0)} +
Q_{nr}^{(0)} q_{rn}^{[1/N] (2)}  \right] + \dots
\end{split}
\end{equation}

Notice that the notation $q_{rn}^{[1/N] (k)}$ refers to the coefficient of the spectral decomposition of the Green's function of order $1/N$ to order $\lambda^k$ in perturbation theory. It is understood that $1+1/N$ has to be such that the series be convergent (this is the case in one and two dimensions).

%We have:
%\begin{equation}
%\begin{split}
%Z^{(0)}\left(1+\frac{1}{N}\right) &= \sum_n \frac{1}{\epsilon_n^{1+1/N}} \\
%Z^{(1)}\left(1+\frac{1}{N}\right) &= \lambda  \left( 1 + \frac{1}{N}\right) \sum_n \frac{\langle n | \sigma | %n\rangle}{\epsilon_n^{1+1/N}} \\
%Z^{(2)}\left(1+\frac{1}{N}\right) &= \frac{\lambda^2}{4} \left(1 + \frac{1}{N} \right) \left[ - \sum_n \frac{ \langle n | \sigma^2 %| n \rangle}{\epsilon_n^{(1+1/N)}}    \right. \nonumber \\
%&+ \left. \frac{1}{2} \sum_{n,m} \frac{(\epsilon_m \epsilon_n)^{-\frac{N+1}{N}} \left(\left(\epsilon _m+3 \epsilon _n\right) %\epsilon_m^{\frac{1}{N}+1}-\left(3 \epsilon_m+\epsilon_n\right) \epsilon_n^{\frac{1}{N}+1}\right)}{\left(\epsilon %_m-\epsilon_n\right)}
%\langle n | \sigma |m \rangle \langle m | \sigma | n \rangle
%\right] \ .
%\end{split}
%\end{equation}

%Observe that the second term in $Z^{(2)}$ is not singular for $m =n$; we can cast the expression for the sum rule in a simpler form  by 
After defining $s= 1+1/N$ we have
\begin{equation}
\begin{split}
Z^{(0)}\left(s\right) &= \sum_n \frac{1}{\epsilon_n^s} \ , \\
Z^{(1)}\left(s\right) &= \lambda  s \sum_n \frac{\langle n | \sigma | n\rangle}{\epsilon_n^s} \ , \\
Z^{(2)}\left(s\right) &= \frac{\lambda^2}{4} s \left[ - \sum_n \frac{ \langle n | \sigma^2 | n \rangle}{\epsilon_n^{s}} \right.  \\
&+ \left. \frac{1}{2} \sum_{n,m} \frac{\left(\left(\epsilon _m+3 \epsilon _n\right) \epsilon_n^{-s}
	-\left(3 \epsilon_m+\epsilon_n\right) \epsilon_m^{-s}\right)}{\left(\epsilon _m-\epsilon_n\right)}
\langle n | \sigma |m \rangle \langle m | \sigma | n \rangle
\right]  \ .
%\\
%&=  \frac{\lambda^2}{4} s \left[ - \sum_n \frac{ \langle n | \sigma^2 | n \rangle}{\epsilon_n^{s}} 
%+ (2s-1) \sum_n \frac{\langle n | \sigma | n\rangle^2}{\epsilon_n^s}
%\right.  \\
%&+ 
%\left. \frac{1}{2} \sum_{n \neq m} \frac{\left(\left(\epsilon _m+3 \epsilon _n\right) \epsilon_n^{-s}
%	-\left(3 \epsilon_m+\epsilon_n\right) \epsilon_m^{-s}\right)}{\left(\epsilon _m-\epsilon_n\right)}
%\langle n | \sigma |m \rangle \langle m | \sigma | n \rangle
%\right]  
\end{split}
\label{eQ_Z1}
\end{equation}

In two dimensions, where the sum rule diverges for $s \leq 1$, yet the expression above allows us to 
approximate the sum rule at $s>1$ arbitrarily close to $s_0=1$ by choosing $N$ large enough.

%Observe that in two dimensions the sum rule diverges as $s \rightarrow 1^+$, i.e. $N \rightarrow \infty$.

\subsection{Order $1/N+1/N'$}

The sum rule of order $1/N +1/N'$ takes the form
\begin{equation}
\begin{split}
Z\left(1+\frac{1}{N}\right) &= \sum_{n,r} q_{nr}^{[1/N]} q_{rn}^{[1/N']}\\
&= \sum_{n,r} q_{nr}^{[1/N] (0)} q_{rn}^{[1/N'] (0)} \\
&+ \lambda \sum_{n,r} \left[ q_{nr}^{[1/N] (0)} q_{rn}^{[1/N'] (1)} +  q_{nr}^{[1/N] (1)} q_{rn}^{[1/N'] (0)}  \right] \\
& + \lambda^2 \sum_{n,r}\left[ q_{nr}^{[1/N] (1)} q_{rn}^{[1/N'] (1)} + q_{nr}^{[1/N] (2)} q_{rn}^{[1/N'] (0)} +
q_{nr}^{[1/N] (0)} q_{rn}^{[1/N'] (2)}  \right] + \dots
\end{split}
\end{equation}
After performing the calculation explicitly we find
\begin{equation}
\begin{split}
Z^{(0)}\left(\frac{1}{N}+\frac{1}{N'}\right) &= \sum_n \frac{1}{\epsilon_n^{1/N+1/N'}} \ , \\
Z^{(1)}\left(\frac{1}{N}+\frac{1}{N'}\right) &= \lambda  \left( \frac{1}{N} + \frac{1}{N'}\right) 
\sum_n \frac{\langle n | \sigma | n\rangle}{\epsilon_n^{1/N+1/N'}} \ , \\
Z^{(2)}\left(\frac{1}{N}+\frac{1}{N'}\right) &= \frac{\lambda^2}{4} \left(\frac{1}{N} + \frac{1}{N'} \right) 
\left[ - \sum_n \frac{ \langle n | \sigma^2 | n \rangle}{\epsilon_n^{(1/N+1/N')}}    \right. \nonumber \\
&+ \left. \frac{1}{2} \sum_{n,m} \frac{(\epsilon_m \epsilon_n)^{-\frac{N+N'}{N N'}} \left(\left(\epsilon _m+3 \epsilon _n\right) \epsilon_m^{\frac{1}{N}+\frac{1}{N'}}-\left(3 \epsilon_m+\epsilon_n\right) \epsilon_n^{\frac{1}{N}+\frac{1}{N'}}\right)}{\left(\epsilon _m-\epsilon_n\right)}
\langle n | \sigma |m \rangle \langle m | \sigma | n \rangle
\right]\ .
\end{split}
\end{equation}

It is remarkable that these expressions take the very same form of the expression  of eq.(\ref{eq_Z1}), after defining 
$s=1/N+1/N'$.

It is easy to verify that the second term in $Z^{(2)}(s)$ is not singular for $m =n$, as it should be; we may cast  the expression for $Z^{(2)}(s)$ in a simpler form by splitting the double series as $\sum_{n,m} = \sum_{n=m} + \sum_{n \neq m}$, i.e.,
\begin{equation}
\begin{split}
Z^{(2)}\left(s\right) 
&=  \frac{\lambda^2}{4} s \left[ - \sum_n \frac{ \langle n | \sigma^2 | n \rangle}{\epsilon_n^{s}} 
+ (2s-1) \sum_n \frac{\langle n | \sigma | n\rangle^2}{\epsilon_n^s}
\right. \nonumber \\
&+ 
\left. \frac{1}{2} \sum_{n \neq m}\frac{\epsilon_m \epsilon_n^{-s} - \epsilon_n \epsilon_m^{-s} +3 \epsilon_n^{1-s} - 3 \epsilon_m^{1-s}}{\left(\epsilon _m-\epsilon_n\right)} 
\langle n | \sigma |m \rangle \langle m | \sigma | n \rangle
\right] \ . \nonumber \\
\end{split}
\end{equation}

Let us now simplify the expression inside the double series:
\begin{equation}
\begin{split}
\frac{\epsilon_m \epsilon_n^{-s} - \epsilon_n \epsilon_m^{-s} +3 \epsilon_n^{1-s} - 3 \epsilon_m^{1-s}}{\left(\epsilon _m-\epsilon_n\right)} &= \left(\epsilon _m^{-s} +\epsilon_n^{-s} \right) + 4 \frac{\epsilon_n^{1-s}-\epsilon_m^{1-s}}{\epsilon _m-\epsilon_n} \ .
\end{split}
\end{equation}

The first contribution can be cast as a single series by suitably renaming indices and using the completeness of the basis:
\begin{equation}
\begin{split}
\sum_{n \neq m} & \left(\epsilon _m^{-s} +  \epsilon_n^{-s} \right) \langle n | \sigma |m \rangle \langle m | \sigma | n \rangle = 2 \sum_{n \neq m} \epsilon_n^{-s}  \langle n | \sigma |m \rangle \langle m | \sigma | n \rangle  \\
&= \sum_n \frac{2}{\epsilon_n^s} \langle n | \sigma \left[ \sum_{m} |m\rangle \langle m | - | n \rangle \langle n | \right] \sigma | n \rangle \\
&= \sum_n \frac{2}{\epsilon_n^s} \left[ \langle n | \sigma^2 | n \rangle - \langle n | \sigma | n \rangle^2 
\right] \ .
\end{split}
\end{equation}

In this way 
\begin{equation}
\begin{split}
Z^{(2)}\left(s\right)  &=  \frac{\lambda^2}{2} s \left[ (s-1) \sum_n \frac{\langle n | \sigma | n\rangle^2}{\epsilon_n^s}
- \sum_{n \neq m}\frac{\epsilon_m^{-s} - \epsilon_n^{1-s}}{\left(\epsilon _m-\epsilon_n\right)} 
\langle n | \sigma |m \rangle \langle m | \sigma | n \rangle
\right]  \ , 
\end{split}
\end{equation}
and the sum rule up to second order reads
\begin{equation}
\begin{split}
Z\left(s\right) &= \sum_n \frac{1}{\epsilon_n^s} \left[
1 + \lambda  s \langle n | \sigma | n\rangle + \frac{\lambda^2}{2} s (s-1)  \langle n | \sigma | n\rangle^2 + \dots
\right] \\
&- \frac{\lambda^2}{2} s \sum_{n \neq m}\frac{\epsilon_m^{-s} - \epsilon_n^{1-s}}{\left(\epsilon _m-\epsilon_n\right)}  \langle n | \sigma |m \rangle \langle m | \sigma | n \rangle + O\left[ \lambda^3\right] \ .
\label{eq:zeta}
\end{split}
\end{equation}

This result agrees with eq.~(9) of Ref.~\cite{Amore12}, which was obtained using the explicit expressions
for the eigenvalues of the operator $\hat{O}$, obtained using perturbation theory~\footnote{Notice that the diagonal
	term in eq.~(9) of Ref.~\cite{Amore12} has been resummed to include higher order contributions.}.

To summarize our findings, we point out that
\begin{itemize}
	\item The complications of Rayleigh-Schr\"odinger perturbation theory when dealing with degenerate states are absent;
	\item The method can be applied systematically to higher order, by calculating the appropriate coefficients of the spectral decomposition of the Green's function of rational order;
	\item For $s \rightarrow 1$ in two dimensions, the sum rules are very sensitive to the asymptotic behavior of the spectrum and therefore they could be used as a tool to study the corrections to Weyl's law;
	\item The case of an operator with a null eigenvalue (not considered here) the traces are ill--defined due to the vanishing
	eigenvalue and a careful renormalization has to be performed; this case will be studied
	separately;
\end{itemize}

\section{Conclusions}
\label{sec:concl}

We have generalized the approach of refs.~\cite{Amore13A,Amore13B,Amore14} to the case of sum rules of rational order $s$.
In doing so, we have introduced Green's functions of order $1/N$, from which the ordinary Green's functions can be
obtained in terms of appropriate convolutions of $N$ of these functions. 
Although we could not obtain an expression for $\tilde{G}^{[1/N]}$ exact to all orders, we have used perturbation theory to obtain 
the explicit expressions up to second order and thus expressed the sum rules of rational order in terms of suitable traces
containing these Green's functions. Our final result of eq.~(\ref{eq:zeta}) reproduces eq.(9) of Ref.~\cite{Amore12} thus confirming 
the findings of that paper, for the case of Dirichlet boundary conditions. The calculation of the analogous sum rules
for the case of a spectrum containing a null eigenvalue, which requires extra care in handling formally divergent contributions,
will be discussed in a separate publication.

\section*{Acknowledgements}
I am grateful to Prof. A.J. Stuart for reading this manuscript and for useful suggestions.
This research was supported by the Sistema Nacional de Investigadores (M\'exico).

\appendix

\begin{appendices}
	
	\section{Spectral decomposition of the Green's function of order $1/2$}
	\label{appA}
	
	\begin{equation}
	\begin{split}
	q_{nm}^{(2)} &= - \frac{1}{8} \Delta_{nm} \langle n | \sigma^2 | m \rangle + \sum_r \frac{1}{4 \eta_{nm}}
	(\Delta_{rr}^2- \Delta_{nr} \Delta_{rm}) \langle n |\sigma | r \rangle \langle r |\sigma | m \rangle  \\
	q_{nm}^{(3)} &= \frac{1}{16} \Delta_{nm} \langle n | \sigma^3 | m \rangle -
	\sum_r \frac{\Delta_{rr}^2}{16 \eta_{nm}} \left(  \langle n | \sigma | r \rangle  \langle r | \sigma^2 | m \rangle 
	+ \langle n | \sigma^2 | r \rangle  \langle r | \sigma | m \rangle  \right) \\
	&- \frac{1}{2\eta_{nm}} \sum_r \left( \Delta_{nr} \langle n | \sigma | r \rangle q_{rm}^{(2)} 
	+ \Delta_{rm} \langle r | \sigma | m \rangle q_{nr}^{(2)}\right) \\
	q_{nm}^{(4)} &= - \frac{5}{128} \Delta_{nm} \langle n | \sigma^4 | m \rangle \\
	&+ \sum_r \frac{\Delta_{rr}^2}{64 \eta_{nm}} \left(  \langle n | \sigma^2 | r \rangle  \langle r | \sigma^2 | m \rangle 
	+ 2 \langle n | \sigma^3 | r \rangle  \langle r | \sigma | m \rangle + 2 \langle n | \sigma | r \rangle  \langle r | \sigma^3 | m \rangle  \right) \\
	&- \sum_r \frac{1}{2 \eta_{nm}} \left( q_{nr}^{(2)} q_{rm}^{(2)} + \Delta_{nr} q_{rm}^{(3)} \langle n |\sigma | r\rangle + \Delta_{rm} q_{nr}^{(3)} \langle r |\sigma | m\rangle \right)  \\
	q_{nm}^{(5)} &= \frac{7}{256} \Delta_{nm} \langle n | \sigma^5 | m \rangle \\
	&- \sum_r \frac{\Delta_{rr}^2}{256 \eta_{nm}} \left(  2 \langle n | \sigma^2 | r \rangle  \langle r | \sigma^3 | m \rangle +
	2 \langle n | \sigma^3 | r \rangle  \langle r | \sigma^2 | m \rangle + 5 \langle n | \sigma | r \rangle  \langle r | \sigma^4 | m \rangle + 5 \langle n | \sigma^4 | r \rangle  \langle r | \sigma | m \rangle \right) \\
	&- \sum_r \frac{1}{2 \eta_{nm}} \left( 2 q_{nr}^{(2)} q_{rm}^{(3)} + 2 q_{nr}^{(3)} q_{rm}^{(2)} + \Delta_{nr} q_{rm}^{(4)} \langle n |\sigma | r\rangle + \Delta_{rm} q_{nr}^{(4)} \langle r |\sigma | m\rangle \right)  \\
	q_{nm}^{(6)} &= -\frac{21}{1024} \Delta_{nm} \langle n | \sigma^6 | m \rangle \\
	&+ \sum_r \frac{\Delta_{rr}^2}{1024 \eta_{nm}} \left[  4 \langle n | \sigma^3 | r \rangle  \langle r | \sigma^3 | m \rangle 
	+ 5 \left( \langle n | \sigma^2 | r \rangle  \langle r | \sigma^4 | m \rangle +  \langle n | \sigma^4 | r \rangle  \langle r | \sigma^2 | m \rangle \right) \nonumber \right. \\
	&+  \left. 14  \left(\langle n | \sigma | r \rangle  \langle r | \sigma^5 | m \rangle 
	+ \langle n | \sigma^5 | r \rangle  \langle r | \sigma | m \rangle \right)\right] \\
	&-  \sum_r \frac{1}{2 \eta_{nm}} \left( 2 q_{nr}^{(3)} q_{rm}^{(3)} + 2 q_{nr}^{(2)} q_{rm}^{(4)}+ 2 q_{nr}^{(4)} q_{rm}^{(2)} + \Delta_{nr} q_{rm}^{(5)} \langle n |\sigma | r\rangle + \Delta_{rm} q_{nr}^{(5)} \langle r |\sigma | m\rangle \right)  \\
	q_{nm}^{(7)} &= \frac{33}{2048} \Delta_{nm} \langle n | \sigma^7 | m \rangle
	- \sum_r \frac{\Delta_{rr}^2}{2048 \eta_{nm}} \left[  5 \left(\langle n | \sigma^3 | r \rangle  \langle r | \sigma^4 | m \rangle +
	\langle n | \sigma^4 | r \rangle  \langle r | \sigma^3 | m \rangle \right) \nonumber \right. \\
	&+ \left. 7 \left( \langle n | \sigma^2 | r \rangle  \langle r | \sigma^5 | m \rangle 
	+  \langle n | \sigma^5 | r \rangle  \langle r | \sigma^2 | m \rangle + 3 \langle n | \sigma | r \rangle  \langle r | \sigma^6 | m \rangle 
	+ 3 \langle n | \sigma^6 | r \rangle  \langle r | \sigma | m \rangle \right) \right] \\
	&-  \sum_r \frac{1}{2 \eta_{nm}} \left[ 2 \left( q_{nr}^{(3)} q_{rm}^{(4)} +  q_{nr}^{(4)} q_{rm}^{(3)}+ q_{nr}^{(2)} q_{rm}^{(5)}+ q_{nr}^{(5)} q_{rm}^{(2)}\right) 
	+ \Delta_{nr} q_{rm}^{(6)} \langle n |\sigma | r\rangle + \Delta_{rm} q_{nr}^{(6)} \langle r |\sigma | m\rangle \right] \\
	q_{nm}^{(8)} &= -\frac{429}{32768} \Delta_{nm} \langle n | \sigma^8 | m \rangle \\
	& +  \sum_r \frac{\Delta_{rr}^2}{16384 \eta_{nm}} \left[ 25 \langle n | \sigma^4 | r \rangle  \langle r | \sigma^4 | m \rangle 
	+ 28 \left(\langle n | \sigma^3 | r \rangle  \langle r | \sigma^5 | m \rangle +  \langle n | \sigma^5 | r \rangle  \langle r | \sigma^3 | m \rangle \right) \nonumber \right. \\
	&+ \left. 42 \left( \langle n | \sigma^2 | r \rangle  \langle r | \sigma^4 | m \rangle +  \langle n | \sigma^4 | r \rangle  \langle r | \sigma^2 | m \rangle \right)
	+ 132 \left( \langle n | \sigma | r \rangle  \langle r | \sigma^7 | m \rangle +  \langle n | \sigma^7 | r \rangle  \langle r | \sigma | m \rangle\right)
	\right] \\
	&-  \sum_r \frac{1}{2 \eta_{nm}} \left[ 2 \left( q_{nr}^{(4)} q_{rm}^{(4)} +  q_{nr}^{(3)} q_{rm}^{(5)}+ q_{nr}^{(5)} q_{rm}^{(3)}+ q_{nr}^{(2)} q_{rm}^{(6)} + + q_{nr}^{(6)} q_{rm}^{(2)}\right) 
	+ \Delta_{nr} q_{rm}^{(7)} \langle n |\sigma | r\rangle + \Delta_{rm} q_{nr}^{(7)} \langle r |\sigma | m\rangle \right] \\
	\end{split}
	\end{equation}

	\section{Values of $\Delta$,$\eta$ and $\xi$ for $N=1,\dots,4$}
	\label{appB}
	
	We report here the values of $\Delta$, $\eta$ and $\xi$ for $N=1, \dots, 4$:
	\begin{table}[h!]
		\begin{center}
			\caption{Expressions of $\Delta_{nm}^{[1/N]}$}
			\vskip .4cm
			\label{tab:table1}
			\begin{tabular}{|c|c|} % <-- Alignments: 1st column left, 2nd middle and 3rd right, with vertical lines in between
				\hline
				$N$ & $\Delta_{nm}^{[1/N]}$ \\
				\hline
%				& \\
				1 &  $\frac{1}{\epsilon_m}+\frac{1}{\epsilon _n}$\\
%				& \\
				2 &  $\frac{\frac{1}{\epsilon_m}+\frac{1}{\epsilon_n}}{\frac{1}{\sqrt{\epsilon_m}}+\frac{1}{\sqrt{\epsilon _n}}}$\\
%				& \\
				3 &  $\frac{\frac{1}{\epsilon_m}+\frac{1}{\epsilon_n}}{\frac{1}{\sqrt[3]{\epsilon_m} \sqrt[3]{\epsilon_n}}+\frac{1}{\epsilon_m^{2/3}}+\frac{1}{\epsilon_n^{2/3}}}$\\
%				& \\
				4 &  $\frac{\frac{1}{\epsilon_m}+\frac{1}{\epsilon_n}}{\frac{1}{\sqrt[4]{\epsilon_m} \sqrt{\epsilon_n}}+\frac{1}{\sqrt{\epsilon _m} \sqrt[4]{\epsilon_n}}+\frac{1}{\epsilon_m^{3/4}}+\frac{1}{\epsilon_n^{3/4}}}$\\	
				\hline
			\end{tabular}
		\end{center}
	\end{table}
	
	\begin{table}[h!]
		\begin{center}
			\caption{Expressions of $\eta_{nm}^{[1/N]}$}
			\vskip .4cm
			\label{tab:table2}
			\begin{tabular}{|c|c|}
				\hline
				$N$ & $\eta_{nm}^{[1/N]}$ \\
				\hline
%				& \\
				1 &  $1$\\
%				& \\
				2 &  $\frac{1}{\sqrt{\epsilon_m}}+\frac{1}{\sqrt{\epsilon _n}}$\\
%				& \\
				3 &  $\frac{1}{\sqrt[3]{\epsilon _m} \sqrt[3]{\epsilon_n}}+\frac{1}{\epsilon_m^{2/3}}+\frac{1}{\epsilon_n^{2/3}}$\\
%				& \\
				4 &  $\frac{1}{\sqrt[4]{\epsilon_m} \sqrt{\epsilon_n}}+\frac{1}{\sqrt{\epsilon_m} \sqrt[4]{\epsilon_n}}+\frac{1}{\epsilon_m^{3/4}}+\frac{1}{\epsilon_n^{3/4}}$\\	
				\hline
			\end{tabular}
		\end{center}
	\end{table}
	
	\begin{table}[h!]
		\begin{center}
			\caption{Expressions of $\xi_{nrm}^{[1/N]}$}
			\vskip .4cm
			\label{tab:table3}
			\begin{tabular}{|c|c|}
				\hline
				$N$ & $\xi_{nrm}^{[1/N]}$ \\
				\hline
%				& \\
				1 &  $0$\\
%				& \\
				2 &  $1$\\
%				& \\
				3 &  $\frac{1}{\sqrt[3]{\epsilon_m}}+\frac{1}{\sqrt[3]{\epsilon_n}}+\frac{1}{\sqrt[3]{\epsilon_r}}$\\
%				& \\
				4 &  $\frac{1}{\sqrt[4]{\epsilon _m}
					\sqrt[4]{\epsilon_n}}+\frac{1}{\sqrt[4]{\epsilon_m} \sqrt[4]{\epsilon_r}}+\frac{1}{\sqrt{\epsilon_m}}+\frac{1}{\sqrt[4]{\epsilon_n} \sqrt[4]{\epsilon_r}}+\frac{1}{\sqrt{\epsilon_n}}+\frac{1}{\sqrt{\epsilon_r}}$\\	
				\hline
			\end{tabular}
		\end{center}
	\end{table}

\end{appendices}

%%%%%%%%%%%%%%%%%%%%%%%%%%%%%%%%%%%%%%%%%%%%%%%%%%%%%%%%%%%%%%%%%%%%%%%%%%%%%%%%%%%%%%%%%%%%%%%%%%%%%%%%%%%%%%%%%%%%%%%%%%%%%%%%%%%%%%%

\end{document}